\documentclass[prl,aps,twocolumn,showpacs]{revtex4}
\usepackage{graphicx}
\usepackage{amsmath}
\begin{document}
%
\title{Superstability of Surface Nanobubbles }
\author{Bram M. Borkent$^{1,2}$, Stephan~M.~Dammer$^{1,2}$,
 Holger Sch\"onherr$^{2,3}$, G. Julius Vancso$^{2,3}$,
and Detlef~Lohse$^{1,2}$}

\affiliation{University of Twente, Faculty of Science and Technology, P.O. Box 217, 7500 AE Enschede, The Netherlands\\
$^1$Physics of Fluids Group, \\
$^2$MESA$^+$ Institute for
Nanotechnology, \\
$^3$Materials Science and Technology of Polymers
Group}

\date{\today}
\begin{abstract}
Shock wave induced cavitation experiments and atomic force
microscopy measurements of flat polyamide and hydrophobized silicon
surfaces immersed in water are performed. It is shown that surface
nanobubbles, present on these surfaces, do {\em not} act as
nucleation sites for cavitation bubbles, in contrast to the
expectation. This implies that surface nanobubbles are not just
stable under ambient conditions but also under enormous reduction of
the liquid pressure down to $-6$MPa. We denote this feature as {\em
  superstability}.
\end{abstract}
\maketitle
In recent years, numerous experiments revealed the existence of
nanoscopic soft domains at the liquid-solid interface,
see~\cite{tyrrellPRL2001,simonsen,holmberga,agrawal,zhangcraig,lijuanzhang,zhang06,jiong,switkes,steitz}
and references therein. Most experiments employ atomic force
microscopy
(AFM)~\cite{tyrrellPRL2001,simonsen,holmberga,agrawal,zhangcraig,lijuanzhang,zhang06,jiong},
but other techniques~\cite{switkes,steitz} have been used as well.
The most consistent interpretation of these experiments is that the
soft domains, which resemble spherical caps with heights of the
order of $10\,{\rm
  nm}$ and diameters of the order of $100\,{\rm nm}$, are so-called {\em
  surface nanobubbles}, i.e., nanoscale gas bubbles located at the
liquid-solid interface. This claim is, for instance, supported by
the fact that nanobubbles can be merged by the tip of an AFM to form
a larger bubble~\cite{simonsen}, or by the fact that they disappear
upon degassing of the liquid~\cite{lijuanzhang,zhang06,switkes},
or by the gas concentration dependence of their density \cite{jiong}.

Surface nanobubbles are puzzling objects. First, {\em they should
not exist}: according to the experimental data these bubbles have a
radius of curvature $R$ of the order of $100\,{\rm nm}$, and
therefore (due to a large Laplace pressure inside of the bubbles)
they should dissolve on timescales far below a
second~\cite{epstein,ljunggren}. In marked contrast
 the
experiments show that nanobubbles are stable for hours. Second, they
are potential candidates to explain various phenomena associated
with the liquid-solid interface, such as liquid slippage at
walls~\cite{lb_pre,lauga,neto} or the anomalous attraction of hydrophobic
surfaces~\cite{tyrrellPRL2001} in water. In addition, 
heterogeneous cavitation usually starts from gaseous
nuclei at solid surfaces (see~\cite{atchley} and references
therein), and surface nanobubbles are suggested as potential
inception sites~\cite{zhang06,bremond05,holmbergb}. However, apart
from convincing experimental evidence for the existence and
stability of nanobubbles, still little is known. For instance, why
are they apparently stable or how do they react to environmental
changes?

In this Letter it is shown that surface nanobubbles, contrary to the
expectation, do {\em not} act as nucleation sites for shock wave
induced cavitation on surfaces, where a large tensile stress is
created in the water. Hence, yet another puzzle is added to the
nanobubble paradox: They are not only stable under ambient
conditions but also under enormous reduction of the water pressure
down to $-6$MPa.
We denote this phenomenon as {\em superstability}.

To demonstrate the superstability of nanobubbles we combine
cavitation experiments and AFM measurements. More precisely,
cavitation experiments (similar
to~\cite{bremond05,bremond06,bremondPOF06}) with different
hydrophobic substrates submerged in water are performed: a shock
wave generates a large tensile stress ($\approx -6\,{\rm MPa}$) in
the water which leads to cavitation of bubbles at the substrates.
The size of the cavitation bubbles is of the order of several
hundred $\mu {\rm m}$. Thus, though the bubbles originate from
smaller nuclei, they can be visualized by optical means. In addition,
AFM-measurements of the same substrates in water at ambient
conditions are performed to proof and quantify the existence of
stable nanobubbles on these substrates. Combining the cavitation and
AFM experiments allows to study the relation between cavitation
activity and nanobubbles. An analogous strategy has been
used previously~\cite{bremond06} to perfectly correlate the
appearance of surface bubbles in cavitation experiments to the
existence of gas-filled microcavities (i.e., microbubbles) of
diameter of $2-4\,\mu{\rm m}$ which had been etched into the surface. Is
there a similar connection between cavitation on smooth unstructured
surfaces and surface nanobubbles?

\begin{figure}[t]
\includegraphics[width=70mm]{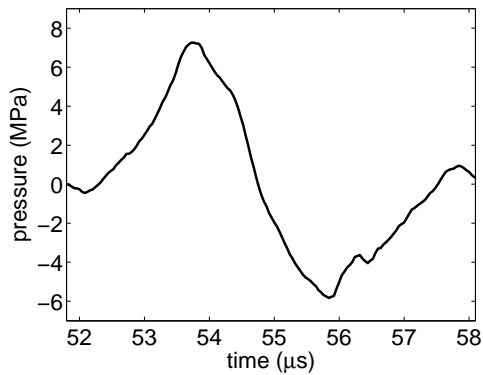}\vspace{0mm}\\
\caption{ \label{pressure} Pressure signal from the shock wave
generator recorded inside of the protective flask with the fibre
optic probe hydrophone close to the surface of the chip. The line
depicts the low pass filtered signal averaged over five recordings.
Triggering the shock wave generator corresponds to time $t=0$.}
\end{figure}

In other words: to what extent must the liquid pressure $p_L$ be
reduced to grow a nanoscale bubble to a visible size
(i.e., above microns)?
A first estimate is obtained from the criterion
that unstable growth of a bubble occurs when $p_L$ falls below the
critical pressure $p_L^c=p_0-p_B$, with the ambient static pressure
$p_0$ and the Blake threshold $p_B$~\cite{brennen95,leighton94}. The
hemispherical dynamics of a surface bubble under rapid decrease of
the liquid pressure is  close to that of a free bubble
with the same radius of curvature~\cite{bremondPOF06}. Therefore,
though surface nanobubbles are spherical caps rather than free
spherical bubbles, one may obtain a reasonable estimate by the
assumption of a spherical bubble. Assuming a nanoscale bubble with
radius $R=100\,{\rm nm}$ and $p_0=1\,{\rm atm}$ one arrives at
$p_L^c\approx -0.55\,{\rm MPa}$ which is exceeded in the experiments
by more than an order of magnitude, see Fig.~\ref{pressure}.
Moreover, we solved the Rayleigh-Plesset
equation~\cite{brennen95,leighton94} (which describes the dynamics
of a spherical bubble under variations of the liquid pressure)
numerically for a gas bubble with the measured
liquid pressure reduction as driving force.
This calculation yields that bubbles down to a
radius of curvature $R=10\,{\rm nm}$ should grow to visible bubbles
during the experiments. Hence, theoretically it should be no problem
to nucleate a surface nanobubble to visible size, but is this
reflected in the experiments?

%

The setup for the cavitation experiments is similar to that used
in~\cite{bremond05,bremond06,bremondPOF06}. A shock wave generator
generates a pressure signal in
the water, consisting of a high pressure front followed by a large
tensile stress, see Fig.~\ref{pressure}.
The substrate of interest is processed and handled inside a filtered
flow bench 
and placed
inside of a sterile flask 
filled with
pure water (Milli-Q Synthesis A10, Millipore), ensuring cleanroom
conditions throughout the experiment. The flask is placed inside the
water tank such that the shock wave is focussed onto the substrate.
The pressure signal at this position is recorded with a fibre optic
probe hydrophone. The cavitation event is photographed by a CCD
camera 
through a long-distance microscope.
The major difference between the present setup
and that of~\cite{bremond05,bremond06,bremondPOF06} is the
maintenance of cleanroom conditions by use of the protective flask.
Compared to Fig.~2 of Ref.~\cite{bremond05} less than 1$\%$
cavitation activity on the surface is observed when cleanroom conditions
 are
preserved, indicating that contaminations play a dominant role for
cavitation experiments under ambient lab conditions.

The AFM data are acquired on a VEECO/Digital Instruments (DI)
multimode AFM equipped with a NanoScope IIIa controller (DI, Santa
Barbara, CA) in tapping mode in water using a DI liquid cell and
V-shaped Si$_3$N$_4$ cantilevers (Nanoprobes, DI). The data shown
for case $D)$ are obtained after mounting the sample into the AFM
while keeping the sample surface covered by water at all times, as
described previously~\cite{morigaki}.

Corresponding to different kinds of substrates and/or different
procedures of substrate preparation, we present results associated
with four different kinds of probes, labeled $A)-D)$. Probes $A)$
and $B)$ use smooth polyamide surfaces as solid substrate. Polyamide
is heated and molded between silicon and atomically smooth mica. The
mica is removed when the polyamide is cooled down to room
temperature, leaving a relatively smooth polyamide surface with a
root mean square (rms) roughness of $3.5\,{\rm nm}$ (measured by AFM
on $1\times1\,\mu{\rm m}^2$) and a static contact angle of
$80^{\circ}$. 
Besides large smooth areas of many ${\rm mm}^2$ the
production process also creates several microscopic cracks in the
surface. 
In case $A)$ these polyamide surfaces are used in the
experiments without further treatment. In case $B)$ the substrate is
first covered by ethanol which is then flushed away by water. This
ethanol-water exchange has been reported to induce the formation of
surface nanobubbles, see~\cite{zhangcraig,jiong} and references therein.
Besides the explanation suggested in~\cite{zhangcraig} we note that
also the exothermic mixing~\cite{exo} of ethanol and water might
induce the formation of nanobubbles, since a temperature increase
favors the formation of nanobubbles~\cite{jiong}. In the cavitation
experiments a drop of ethanol is placed on the substrate such that
it is completely covered by ethanol before it is submerged in water.
Then the substrate is moved inside the protective flask for a
minute to replace the miscible ethanol by water~\cite{comment}. In
the AFM experiment for case $B)$ a liquid cell is used.

Probes $C)$ and $D)$ use pieces of smooth hydrophobized silicon as
the substrate. A Si(100) wafer is diced into chips ($1\times1\,{\rm
cm}^2$) which are immersed for 15 minutes in a (5:1) Piranha
cleaning mixture. Hereafter, the chips are hydrophobized by chemical
vapor deposition of 1H,1H,2H,2H-perfluorodecyldimethylchlorosilane
(PFDCS)~\cite{bremond05}, yielding a rms value of $\approx
0.36\,{\rm nm}$ (measured by AFM on $1\times1\,\mu{\rm m}^2$), a
coating thickness $\approx 2.6\,{\rm nm}$ (measured by
ellipsometry), and an advancing contact angle $\approx 100^\circ$.
Note that the silane-film is not able to move; it is a stable 
self-assembled monolayer covalently bonded to the underlying substrate.
Before immersion in water the probes are rinsed with ethanol and
blown dry with argon gas~\cite{bremond05}. Case $D)$ additionally
applies the \emph{ex situ} ethanol-water exchange as described
above.

In each of the cases $A)-C)$ substrates of the respective type are
produced in an identical manner. One substrate is used in the cavitation
experiments and one in the AFM measurements. Note that we checked that the
observed cavitation activity and nanobubble density were
reproducible among substrates of the same kind. Furthermore, 
in case $D)$ the \emph{same} substrate is
used in both experiments. After the cavitation experiment (exposure
to a single shock wave) the substrate is transported in water to the
AFM, where it is mounted without exposure to air, whereafter the
water-solid interface is imaged as it appears after the cavitation
experiments.

\begin{figure*}[t]
\includegraphics[width=140mm]{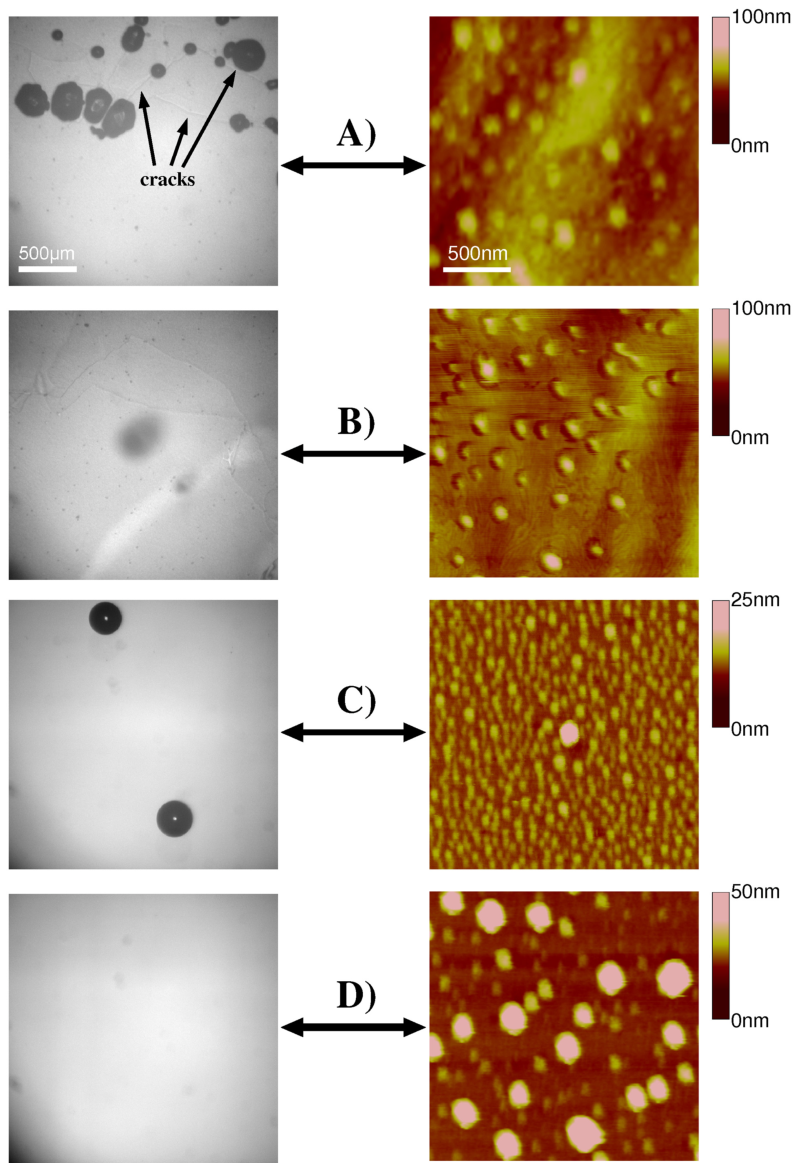}\vspace{0mm}\\
\caption{ \label{result}(color online) Cavitation activity (left),
and corresponding nanobubble density (right) imaged by AFM
(topography images) for various probes. 
The length scales given in A) also refer to B), C), and D). --
$A)$ and $B)$: polyamide
substrates, $B)$ after ethanol-water exchange (see text). $C)$ and
$D)$: hydrophobized silicon substrates, $D)$ after ethanol-water
exchange. Scanning velocities for the cases $A)-D)$ are 46, 20, 10,
and 40 $\mu \rm{m /s}$ respectively. There is hardly any cavitation
though the substrates are densely covered by surface nanobubbles.
Note that the cavitation bubbles in $A)$ emerge exclusively from
microscopic cracks in the surface, whereas the whole substrate is
covered by nanobubbles. The cavitation bubbles in $C)$ presumably
originate from surface contaminations. In $D)$ it is {\em
explicitly} shown that the nanobubbles are still stably present {\em
after} the cavitation experiments. 
}
\end{figure*}

Do substrates with a high nanobubble density show a large cavitation activity?
Fig.~\ref{result} illustrates the experimental results.
The left panel
shows typical recordings of the
cavitation experiments for the cases $A)-D)$. The right panel shows
the corresponding AFM-measurements of the substrate surfaces
immersed in water. Though the substrates have relatively large
contact angles and the water pressure drops down to $\approx
-6\,{\rm MPa}$ during the experiments there is hardly any
cavitation on
the smooth substrates $A)-D)$. Note that the cavitation bubbles in
$A)$ originate exclusively from microscopic cracks in the surface,
as can be seen in Fig.~\ref{result}A). Applying the ethanol-water
exchange, these microcracks do not lead to surface cavitation, see
Fig.~\ref{result}B). Contrary to the cavitation experiments, the AFM
measurements show that all substrates are densely covered by surface
nanobubbles, with number densities between 10 and 80 bubbles per
$\mu \rm m^2$. The sizes range from 3 to 40 nm in height and 60 to
300 nm in diameter. Several standard tests were performed (not
shown) to ensure that the structures seen in the AFM images are
indeed surface nanobubbles. Force-volume
measurements~\cite{tyrrellPRL2001,holmberga,zhangcraig} and tip
manipulation of the bubbles~\cite{simonsen} are in accordance with
previous studies. Furthermore, nanobubbles are not present when the
substrates are immersed in ethanol, in agreement with~\cite{jiong}.
Successive cycles of ethanol-water and water-ethanol exchange
resulted in pictures without (in ethanol) and with nanobubbles (in
water). Finally, when degassed ethanol is exchanged by degassed
water, nanobubbles are \emph{not} induced.

Thus the combination of the cavitation and the AFM experiments yields the
remarkable result that the surface nanobubbles do not cavitate,
in spite of the enormous tensile stress they
are exposed to. This contradicts the expectation
that the experimental pressure signal should be able to
cavitate bubbles with an initial radius of curvature down to 8\,nm.
Case $D)$ explicitly shows that nanobubbles are still present {\em
after} the cavitation experiments, and that there is no cavitation
activity at the surface induced by the shockwave.
While it is already puzzling that surface nanobubbles
are stable under ambient conditions, it is even more puzzling that
they still exist after the passage of a shock wave with a large
tensile stress down to $\approx -6\,{\rm
  MPa}$. We denote this as {\em superstability}.

One may wonder what actually is happening with the surface nanobubbles
when the shock wave is passing by.
With the present  technology it 
is impossible to AFM-image the
nanobubbles  (which takes order of minutes) during the
shock wave passage (which is order of $\mu s$).
Therefore, evidence can only be indirect.

One may also question whether the nanobubbles survive the compression
wave (with typical time scale $\tau \approx 1 \mu s$ so that the nanobubbles
respond quasi-statically).
During the compression phase,
gas may diffuse into the neighboring
   liquid around the bubble. 
With a typical diffusion constant of $D \approx 10^{-9} m^2/s$ 
we get as typical diffusion length scale $\sqrt{\tau D} \approx 100 nm$.
Hence the liquid close to the remaining void
   (100nm) will become supersaturated with gas. 
However, during the negative pressure phase, i.e., during the 
expansion of the bubble, all this gas will be recollected by the 
bubble, as has been shown in 
ref.\ \cite{hil96} (for micrometer bubbles).

In summary, it is demonstrated that in standard shock wave induced
cavitation experiments surface nanobubbles do {\em not} act as
nucleation sites. Cavitation bubbles originate from contaminations
or from microscopic structures such as microcracks or
microcrevices~\cite{bremond06,bremondPOF06}, rather than from
surface nanobubbles which densely populate the immersed substrates.
This implies that surface nanobubbles are unexpectedly stable
under large tensile stresses.

We thank Szczepan Zapotoczny for his contribution,
K.\ A.\ Morch for discussions, and acknowledge
financial support from STW
(NanoNed Program), DFG
(Grant No. DA969/1-1)), the MESA$^{+}$ Institute,
and 
CW-NWO 
(vernieuwingsimpuls program  for H.S.).


\begin{thebibliography}{99}
\bibitem{tyrrellPRL2001} J.~W.~G.~Tyrrell and P.~Attard, Phys.~Rev.~Lett. {\bf
    87}, 176104 (2001).
\bibitem{simonsen} A.~C.~Simonsen, P.~L.~Hansen, and B.~Kl\"osgen, J. Colloid
  Interface Sci. {\bf 273}, 291 (2004).
\bibitem{holmberga} M.~Holmberg {\em et al.}, Langmuir {\bf 19}, 10510 (2003).
\bibitem{agrawal} A.~Agrawal {\em et al.}, Nano Lett.~{\bf 5}, 1751 (2005).
\bibitem{zhangcraig} X.~H.~Zhang, N.~Maeda, and V.~S.~J.~Craig, Langmuir {\bf 22},
  5025 (2006).
\bibitem{lijuanzhang} L.~Zhang {\em et al.}, Langmuir {\bf 22}, 8109 (2006).
\bibitem{zhang06} X.~H.~Zhang {\em et al.}, Langmuir \textbf{22},
9238 (2006).
\bibitem{jiong} S.~Yang {\em et al.}, submitted.
\bibitem{switkes} M.~Switkes and J.~W.~Ruberti, Appl.~Phys.~Lett.~{\bf 84}, 4759
  (2004).
\bibitem{steitz} R.~Steitz {\em et al.}, Langmuir {\bf 19}, 2409
  (2003).
\bibitem{epstein} P.~S.~Epstein and M.~S.~Plesset, J.~Chem.~Phys.~{\bf 18}, 1505
  (1950).

\bibitem{ljunggren} S.~Ljunggren and J.~C.~Eriksson, Colloids~Surf.~A {\bf 129},
  151 (1997).
\bibitem{lb_pre} E.~Lauga and M.~P.~Brenner, Phys. Rev. E {\bf 70},
026311 (2004).
\bibitem{lauga} E.~Lauga, M.~P.~Brenner, and H.~A.~Stone, in {\em Handbook of
    Experimental Fluid Dynamics}, edited by C.~Tropea, J.~Foss, and A.~Yarin
    (Springer, New York 2005).
\bibitem{neto} C.~Neto {\em et al.}, Rep.~Prog.~Phys. {\bf 68}, 2859 (2005).
\bibitem{atchley} A.~A.~Atchley and A.~Prosperetti, J.~Acoust.~Soc.~Am.~{\bf 86},
1065 (1989).
\bibitem{holmbergb} M.~Holmberg {\em et al.}, in: Fifth International Symposium on Cavitation
(2003).
\bibitem{bremond05}
N.~Bremond, M. Arora, C. D. Ohl and D. Lohse,
J. Phys.:~Condens.~Matter {\bf 17},
  S3603 (2005).
\bibitem{bremondPOF06}
N.~Bremond, M. Arora, C. D. Ohl and D. Lohse,
Phys. Fluids.
\textbf{18}, 121505 (2006).
\bibitem{bremond06}
N.~Bremond, M. Arora, C. D. Ohl and D. Lohse,
Phys.~Rev.~Lett.~{\bf 96}, 224501
  (2006).
\bibitem{morigaki} K.~Morigaki~{\em et al.}, Langmuir
\textbf{19}, 6994 (2003).
\bibitem{brennen95} C.~E.~Brennen, {\em Cavitation and Bubble Dynamics} (Oxford
  University Press, New York, 1995).
\bibitem{leighton94} T.~G.~Leighton, {\em The Acoustic Bubble} (Cambridge
  University Press, Cambridge, 1994).


\bibitem{exo} {\em Rodd's chemistry of carbon compounds}, 2nd edition, vol 1,
  part B, edited by S.~Coffey (Elsevier, Amsterdam, 1965).

  \bibitem{comment}We verified that our \emph{ex situ} ethanol-water exchange generates surface nanobubbles, by placing the processed
substrate in the AFM while keeping water on its surface. Then the
substrate was AFM-scanned and many surface nanobubbles ($>10$ per $\mu
{\rm m}^2$) were observed.

\bibitem{hil96}
S. Hilgenfeldt, D.  Lohse, and M. P. Brenner,
Phys. Fluids {\bf 8}, 2808 (1996).

\end{thebibliography}
\end{document}